\documentclass[aps,twocolumn,showpacs,preprintnumbers,amsmath,amssymb,nofootinbib,superscriptaddress,showkeys]{revtex4}

\usepackage{epsfig}
\usepackage{graphicx}

\begin{document}

\title{A Two Potential Formula and its Application to Proton-Proton Scattering}
\author{M. Pav\'on Valderrama}\email{m.pavon.valderrama@fz-juelich.de}
\affiliation{Institut f\"ur Kernphysik and J\"ulich Center for Hadron Physics, Forschungszentrum J\"ulich, 52425 J\"ulich, Germany} 
\author{E. Ruiz Arriola}\email{earriola@ugr.es}
\affiliation{Departamento de F\'{\i}sica At\'{o}mica, Molecular y Nuclear, Universidad de Granada, E-18071 Granada, Spain.}

\date{\today}

\begin{abstract} 
\rule{0ex}{3ex}
Within the framework of potential scattering theory we derive an analytical
two-potential formula for the on-shell partial wave scattering amplitude.
This formula embodies a large number of possible applications,
including long range Coulomb forces as well as short distance singular
potentials.
As an example illustrating the use of the formula we
analyze the determination of the strong proton-proton scattering
s-wave phase shift from the experimentally determined Coulomb phase
when the one pion exchange and two pion exchange chiral potentials are
taken into account and analyze the relevant scales of the problem.

\end{abstract}


\pacs{03.65.Nk,11.10.Gh,13.75.Cs,21.30.Fe,21.45.+v}
\keywords{Scattering theory,Proton-proton interaction, One and Two
  Pion Exchange,Renormalization.}

\maketitle 

\section{Introduction}
\label{sec:intro}

The two-potential formalism, developed in the fifties by 
Watson~\cite{PhysRev.88.1163} and
Gell-Mann and Goldberger~\cite{PhysRev.91.398},
relates the scattering due to the sum of two different potentials and
has a widespread use in scattering theory.
The usual example is the treatment of Coulomb distortion for strongly
interacting particles.
The problem is to determine the total scattering amplitude $T$ from a
potential constructed as the sum of two potentials $V = V_S + V_L$ 
in terms of the scattering amplitude $T_S$ due only to the potential $V_S$.
The result can be found in a straightforward manner
using the Lippmann-Schwinger equation
\begin{eqnarray} 
T &=& V + V G_0 T \, ,
\end{eqnarray} 
with $V$ the potential operator and $G_0 = (E-H_0 )^{-1} $ the resolvent
of the free Hamiltonian.
The outgoing boundary condition corresponds to $ E \to E + {\rm i} 0^+ $.
The $T$-matrix can then be expressed as~\footnote{We have chosen this
particular formulation of the two-potential trick in order to have
a more consistent notation along the paper.
Normally it is written in terms of the distorted
short range $T$-matrix and the undistorted long range $T$-matrix,
but in either case the resulting $T$-matrix is the same.
}
\begin{eqnarray} 
\label{eq:two-potential-trick}
T= T_S + (1 + T_S G_0)\,\hat{T}_L\,(G_0 T_S + 1) \, ,
\end{eqnarray} 
where $T_S$ is the short distance $T$-matrix, and $\hat{T}_L$ the long 
distance one distorted by short range effects
\begin{eqnarray} 
T_S &=& V_S + V_S G_0 T_S \, , \\
\hat{T}_L &=& V_L + V_L G_S \hat{T}_L \, ,
\end{eqnarray} 
with $G_S = G_0 + G_0 V_S G_S$ the full propagator for $V_S$.

While the result above solves the problem, it does not explicitly relate the
on-shell scattering amplitudes of the full and short distance potentials.
The reason is that the Lippmann-Schwinger equation involves
the off-shell behaviour of the potentials,
which allows to treat nonlocal potentials quite
straightforwardly, but precisely because of this it is hard to profit
specifically from the simplifying features which arise in the
interesting and quite frequent case of local potentials arising
e.g. in a particle exchange picture.
For the local case, a coordinate space formulation of the scattering problem
is more convenient (an effective field theory example is provided by 
Ref.~\cite{Entem:2007jg}).

In this paper we derive a two-potential formula, which relates the
phase shifts (i.e. the on-shell scattering matrix) of the full and
short range potentials $V$ and $V_S$, and which is based on two
assumptions: (i) the potentials are local and (ii) the short range
potential dominates at short distances.
Our result will be amenable to rather detailed analytical study, hence
enlarging the class of situations one may cover. The connection to
momentum space renormalization with counterterms of the
Lippmann-Schwinger equation is also analyzed.  This is particularly
enlightening in the case of singular potentials and their
renormalization, a subject of recent interest (see
e.g. \cite{Entem:2007jg}).  As an illustrative application
we show how our two-potential formula may be used to deduce the strong
proton-proton phase shifts from the experimentally measured ones when
long distance corrections coming from one and two pion exchange
contributions are taken into account.

\section{Two potential formula}

We consider the non-relativistic scattering of two particles by a spherically
symmetric potential $V$ which can be decomposed into the following two pieces
\begin{eqnarray}
V(r) = V_L(r) + V_S(r) \, ,
\end{eqnarray}
where $V_L$ and $V_S$ respectively represent the long and short distance
components of the interaction.
We will assume that the short range potential $V_S$ dominates at
short distances $r = r_c$, i.e.
\begin{eqnarray}
\label{eq:condition-short}
V_S(r_c) \gg V_L(r_c) \, ,
\end{eqnarray}
for $r_c$ small enough.
The system can be described by solving the reduced Schr\"odinger equation
(for simplicity, we only consider here the s-wave case)
\begin{eqnarray}
\label{eq:schroe-full}
-u_k'' + 2\mu\,\left[V_L(r) + V_S(r)\right]\,u_k(r) = k^2 u_k(r) \, ,
\end{eqnarray}
with $u_k(r)$ the reduced wave function, $\mu$ the reduced mass of the
two body system and $k$ the center-of-mass momentum.
The phase shift can then be obtained by matching the reduced wave 
function $u_k$ to the usual asymptotic boundary condition for $r\to\infty$
\begin{eqnarray}
\label{eq:ps-uk}
u_k(r) \to \cot{\delta}\,\sin{k r} + \cos{k r} \, ,
\end{eqnarray}
where we have assumed that the long range potential $V_L$ decays faster than
$1/r^2$ at large distances, so phase shifts are well defined.
We also consider the corresponding scattering problem for which only the
short range potential $V_S$ is present.
In such a case, the reduced Schr\"odinger equations reads
\begin{eqnarray}
\label{eq:schroe-short}
-{u_k^S}'' + 2\mu\,V_S(r)\,u_k^S(r) = k^2 u_k^S(r) \, ,
\end{eqnarray}
with $u_k^S$ the {\it short} reduced wave function.
The phase shift can be extracted from the asymptotic behaviour of $u_k^S$
\begin{eqnarray}
u_k^S(r) \to \cot{\delta^S}\,\sin{k r} + \cos{k r} \, ,
\end{eqnarray}
for $r \to \infty$.

The problem is to relate the full phase shift $\delta(k)$ with the 
short phase shift $\delta^S(k)$.
For that purpose, we will assume that at short enough distances $r =
r_c$ the full and short reduced wave function $u_k$ and $u_k^S$ are
approximately equal, $u_k(r_c) \simeq u_k^S(r_c)$.
The previous approximation can be restated in terms of the logarithmic
derivatives of the reduced wave functions~\footnote{We are not necessarily
assuming a regular solution of the Schr\"odinger equation, i.e. $V_S$ can
contain zero-range pieces.}
\begin{eqnarray}
\label{eq:bc-equality}
\frac{u_k'(r_c)}{u_k(r_c)} = \frac{{u_k^S}'(r_c)}{u_k^S(r_c)} \, .
\end{eqnarray}
This expression will hold true when the condition expressed
in Eq.~(\ref{eq:condition-short}) is fulfilled.
We now make use of the superposition principle to represent the full
solution as the following linear combination
\begin{eqnarray}
\label{eq:u-lin}
u_k(r) &=& \cot{\delta}\,J_k(r) - Y_k(r) \, ,
\end{eqnarray}
where $J_k$ and $Y_k$ are solutions of Eq.~(\ref{eq:schroe-full}), subjected
to the asymptotic conditions
\begin{eqnarray}
J_k(r) &\to&  \sin{k r} \, , \\ Y_k(r) &\to& - \cos{k r} \, .
\end{eqnarray}
Analogously, we write the short range solution as
\begin{eqnarray}
\label{eq:uS-lin}
u_k^S(r) = \cot{\delta^S}\,J_k^S(r) - Y_k^S(r) \, ,
\end{eqnarray}
with $J_k^S$ and $Y_k^S$ solutions of Eq.~(\ref{eq:schroe-short}), such that
\begin{eqnarray}
J_k^S(r) &\to& \sin{k r} \, , \\
Y_k^S(r) &\to& - \cos{k r} \, ,
\end{eqnarray}
for $r \to \infty$.
By matching the logarithmic derivatives we arrive at our final expression
\begin{eqnarray}
\label{eq:2V-formula}
\cot \delta (k) = 
\frac{{\cal A} (k,r_c) \cot \delta^S (k) - {\cal B} (k,r_c) }
{{\cal C} (k,r_c) \cot \delta^S (k) - {\cal D} (k,r_c) } \, ,
\end{eqnarray} 
where $\cal A$, $\cal B$, $\cal C$ and $\cal D$ are defined as
\begin{eqnarray}
{\cal A} (k,r_c) &=& W(J_k^S, Y_k)\big|_{r = r_c} \, \\
{\cal B} (k,r_c) &=& W(Y_k^S, Y_k)\big|_{r = r_c} \, \\
{\cal C} (k,r_c) &=& W(J_k^S, J_k)\big|_{r = r_c} \, \\
{\cal D} (k,r_c) &=& W(Y_k^S, J_k)\big|_{r = r_c} \,
\end{eqnarray} 
with $W(f, g)|_{r = r_c} = f(r_c) g'(r_c) - f'(r_c) g(r_c)$ the Wronskian
between different wave functions evaluated at the cut-off radius $r = r_c$.
The bilinear structure is reminiscent of the Moebius transformation
invariance discussed at length in Ref.~\cite{PavonValderrama:2007nu}
in the context of the renormalization group analysis
with boundary conditions.
It should be noted that although the matching of log-derivatives in
order to obtain long range correlations is not new, one nice example
being the Landau-Smorodinsky derivation of the effective range
expansion~\cite{Landau44}, or the treatment of hadronic atoms in
Ref.~\cite{Holstein:1999nq}, its use in combination with the
superposition principle in order to derive direct relations between
phase shifts is less common, and it has only been partially exploited
in some effective field theory r-space
computations~\cite{PavonValderrama:2005gu,Valderrama:2005wv,cordon:054002}. 

In passing we also note that Eq.~(\ref{eq:2V-formula}) cannot be
derived from the Lippmann-Schwinger equation.
The reason is that the two-potential formula depends on the {\it explicit}
formulation of the following: (i) the superposition principle via
Eqs.~(\ref{eq:u-lin}) and (\ref{eq:uS-lin}), and  (ii) the short
distance boundary condition for the Schr\"odinger equation, 
Eq.~(\ref{eq:bc-equality}).
These two conditions are included in the Lippmann-Schwinger equation, 
but {\it implicitly}, in a way that they cannot be directly handled, 
impeding the derivation of the previous formula (but allowing
the derivations of formulas relating the full off-shell
scattering amplitudes, like the two-potential trick,
Eq.~(\ref{eq:two-potential-trick})).

The two potential formula also holds in certain cases for non-local potentials.
The necessary condition for its application is that the non-local potential
does not involve derivatives of order higher than two, e.g. potentials of
the type 
\begin{eqnarray}
V_S^{NL} = \{\nabla^2, f_S(r)\} \, ,
\end{eqnarray}
where $\{,\}$ represents the anti-commutator.
In such a case the short distance boundary condition
for the Schr\"odinger equation can be expressed as the
log-derivative of the wave function, Eq.~(\ref{eq:bc-equality}).
For non-local potentials involving higher derivatives, the two-potential
formula can still be applied when the cut-off radius $r_c$ is larger than
the range at which the non-localities appear.

\section{Coulomb scattering}

The case where the long range potential is of Coulomb type requires a
special treatment as the usual asymptotic behaviour described 
in Eq.~(\ref{eq:ps-uk}) does not apply.
For definiteness, we analyze here the Coulomb repulsion between two
unit charge particles in the s-wave.
The full system is now described by the following equation
\begin{eqnarray}
\label{eq:schroe-coulomb-full}
-{u_k^C}'' + 2\mu\,\left[V_S(r) + \frac{\alpha}{r}\right]\,u_k^C(r) &=& 
k^2 u_k^C(r) \, ,
\end{eqnarray}
where we have added the ${}^C$ superscript for labelling the Coulomb solution,
and $\alpha$ represents the fine structure constant.
The correct asymptotics for $u_k^C$ is given by
\begin{eqnarray}
u_k^C(r) &\to& \cot{\delta^C}\,F_0(\eta,\rho) + G_0(\eta,\rho) \, ,
\end{eqnarray}
with $\delta^C$ the Coulomb modified phase shift, and
$F_0(\eta,\rho)$ and $G_0(\eta,\rho)$ the usual s-wave Coulomb wave
functions (see for example~\cite{abramowitz+stegun}), which depend on the
parameters $\eta = 1/(k a_B)$ and $\rho = k r$; $a_B = 1 / (\mu \alpha)$
is the Bohr radius of the two particle system.
$F_0$ and $G_0$ are solutions of the reduced Schr\"odinger equation 
for the Coulomb potential $V_C(r) = \alpha/r$, with the asymptotic
behaviour
\begin{eqnarray}
F_0(\eta, \rho) &\to& \sin{(k r - \eta \log{2 k r} + \sigma_0)} \, , \\
G_0(\eta, \rho) &\to& \cos{(k r - \eta \log{2 k r} + \sigma_0)} \, ,
\end{eqnarray}
where $\sigma_0$ is a phase shift defined as 
$\sigma_0 = {\rm arg}\,\Gamma(1+i\eta)$.
As in the previous case, we can use the superposition principle to rewrite
the full solution
\begin{eqnarray}
u_k^C(r) &=& \cot{\delta^C}\,F^C_k(r) - G^C_k(r) \, ,
\end{eqnarray}
with $F_k(r)$ and $G_k(r)$ solutions of Eq.~(\ref{eq:schroe-coulomb-full})
subjected to the asymptotic boundary conditions
\begin{eqnarray}
F^C_k(r) &\to& F_0(\eta,\rho) \, , \\
G^C_k(r) &\to& - G_0(\eta,\rho) \, .
\end{eqnarray}
The short range system is described by Eq.~(\ref{eq:schroe-short}),
and the short range wave function $u_k^S$ is again
parametrized by Eq.~(\ref{eq:uS-lin}).

After matching logarithmic derivatives we find
\begin{eqnarray}
\label{eq:2V-formula-coulomb}
\cot \delta^C (k) = 
\frac{{\cal A} (k,r_c) \cot \delta^S (k) - {\cal B} (k,r_c) }
{{\cal C} (k,r_c) \cot \delta^S (k) - {\cal D} (k,r_c) }
\end{eqnarray} 
where $\cal A$, $\cal B$, $\cal C$ and $\cal D$ are now defined as
\begin{eqnarray}
{\cal A} (k,r_c) &=& W(J_k^S, G_k^C)\big|_{r = r_c} \, \\
{\cal B} (k,r_c) &=& W(Y_k^S, G_k^C)\big|_{r = r_c} \, \\
{\cal C} (k,r_c) &=& W(J_k^S, F_k^C)\big|_{r = r_c} \, \\
{\cal D} (k,r_c) &=& W(Y_k^S, F_k^C)\big|_{r = r_c} \,
\end{eqnarray}
in complete analogy with Eq.~(\ref{eq:2V-formula}).
Previous relationships between Coulomb and short distance scattering can
be found for some specific cases
in Refs.~\cite{Partensky1967382,Kok:1982dr,deMaag:1983nz}.

\subsection{Contact Short Range Potential}

A simple application of the previous formula corresponds to a
situation where the short range potential is zero for distances greater
than the cut-off radius $r_c$
\begin{eqnarray}
V_S(r) = 0 \quad \mbox{for $r > r_c$} \, ,
\end{eqnarray}
while, for distances shorter than $r_c$, the potential is very strong.
The previous potential corresponds to a $\delta$-type contact interaction
regularized at the length scale $r_c$.
In such a case, the $F_k^C(r)$ and $G_k^C(r)$ wave functions are equal to
their asymptotic behaviour for $r \geq r_c$, and by taking into account
their behaviour at small radii
\begin{eqnarray}
F_k^C(r) &\to& k\,C(\eta)\,\Big[ r + \frac{r^2}{a_B} + O(r^3) \Big] \, ,\\
G_k^C(r) &\to& - \frac{1}{C(\eta)}\,\Big[ 1 + \frac{2 r}{a_B}\,
\left( \log{\frac{2 r}{a_B}} + 2\,\gamma_E - 1 + h(\eta) \right) \nonumber \\
&&  + O(r^2) \Big] \, ,
\end{eqnarray}
with $\gamma_E$ the Euler-Mascheroni constant, and 
$C(\eta)$ and $h(\eta)$ defined as
\begin{eqnarray}
C^2(\eta) &=& \frac{2 \pi \eta}{e^{2 \pi \eta} - 1} \, , \\
h(\eta) &=& \eta^2\,\sum_{n=1}^{\infty}\,\frac{1}{n (n^2 + \eta^2)} - 
\log{\eta} - \gamma_E \, ,
\end{eqnarray}
the relationship given by Eq.~(\ref{eq:2V-formula-coulomb}) can be evaluated
explicitly, yielding
\begin{eqnarray}
\label{eq:kcotd-rel-unreg}
k\,\cot{\delta^S(k)} &=& {C}^2(\eta) k\,\cot{\delta^C(k)} + 
2 \frac{h(\eta)}{a_B} \nonumber \\ &&
-  \frac{2}{a_B}\Big[  \log{\frac{a_B}{2 r_c}} - 2\gamma_E \Big] + 
{\mathcal O}(r_c) \, ,
\end{eqnarray}
where terms linear in the cut-off radius and higher powers of $r_c$
have been ignored.
As can be seen the previous expression is logarithmic divergent with respect
to $r_c$, but can be regularized if we take into account the corresponding
expression for $k = 0$, which is similar to the well known 
relationship between strong and Coulomb scattering
length from Blatt and Jackson~\cite{Blatt49,RevModPhys.22.77}
\begin{eqnarray}
\label{eq:aS-aC-contact}
- \frac{1}{\alpha_{S}} = - \frac{1}{\alpha_{C}} - 
\frac{2}{a_B}\left[ \log{\frac{a_B}{2 r_c}} - 2\,\gamma_E \right] +
{\mathcal O}(r_c) \, .
\end{eqnarray}
The expression above diverges in exactly
the same way as Eq.~(\ref{eq:kcotd-rel-unreg}).
Subtracting the $k=0$ expression to Eq.~(\ref{eq:kcotd-rel-unreg}), and
taking the $r_c \to 0$ limit, we arrive at the following expression
\begin{eqnarray}\label{eq:coulomb-contact}
k\,\cot{\delta^S(k)} + \frac{1}{\alpha_{S}} &=& 
{C}^2(\eta) k\,\cot{\delta^C(k)} \nonumber \\ && 
+ 2\,\frac{h(\eta)}{a_B} + \frac{1}{\alpha_{C}} \, . 
\end{eqnarray}
The expected error of this formula can be estimated reintroducing
the cut-off $r_c$ and interpreting it as the neglected range $R_S$
of the short range potential $V_S$, yielding a relative error 
of ${\cal O}( r_c / a_B) = {\cal O} (R_S / a_B)$.

The corresponding formula for attractive Coulomb interaction 
may be of interest for the treatment of pionic atoms, 
and can be obtained by taking $\eta = - 1/(k a_B)$ negative,
and making the following substitution
\begin{eqnarray}
h(\eta) &\to&
{\rm Re}\left[ \psi(i\eta) - \log{(-i\eta)}\right]
\end{eqnarray}
with $\psi$ the digamma function.

Finally, the corresponding formula for p-wave repulsive Coulomb interaction
can be worked out analogously to the s-wave case. 
The treatment of the divergences is nonetheless more involved: there is an
additional logarithmic divergence proportional to $k^2$, due to the
interplay between the Coulomb potential and the centrifugal barrier.
The outcome is that two subtractions are needed in order to have finite
results in the $r_c \to 0$ limit.
The final formula is rather simple to summarize
\begin{eqnarray}\label{eq:coulomb-contact-p-wave}
k^3\,\cot{\delta^S_1(k)} &+& \frac{1}{\alpha_{1,S}} - \frac{1}{2} r_{1,S} k^2
=  {C_1^2}(\eta)\,k^3\cot{\delta_1^C(k)} \nonumber\\
&+& k^3\,(1 + \eta^2)\,2\,\eta\,h(\eta)
\nonumber \\ &+& 
\frac{1}{\alpha_{1,C}} - \frac{1}{2} r_{1,C} k^2 \, ,
\end{eqnarray}
where $C_1^2(\eta) = (1+\eta^2) C^2(\eta)$, $\alpha_{1,S}$
and $\alpha_{1,C}$ are the p-wave short and Coulomb scattering
volumes, and $r_{1,S}$ and $r_{1,C}$ the p-wave short and Coulomb
effective ranges.
The previous formula has less predictive power than the corresponding one 
for s-waves as a consequence of the extra subtraction needed
to regularize it.
A possible application is nucleon-alpha
scattering~\cite{Bertulani:2002sz}.

\section{Application to Proton-Proton Scattering}

Now we apply the previous results for Coulomb scattering to the
specific case of proton-proton (pp) scattering in s-waves.
We consider the strong pp interaction as the short range
potential $V_S$, while the Coulomb repulsion between the protons
plays the role of the long range potential $V_L$.

\subsection{Pionless theory}
\label{subsec:pionless}

We first consider the simplifying case in which the pion exchange
interactions between the protons are neglected and the pp potential
consists on contact interactions only, i.e. the pionless theory,
characterized by a short distance boundary condition.
In such a case the two potential formula given by
Eq.~(\ref{eq:coulomb-contact}) applies.
The previously mentioned relationship can be better understood by noticing
the relationship with the strong and Coulomb effective range
expansions~\cite{PhysRev.76.38}, i.e.
\begin{eqnarray}
k\,\cot{\delta^S} &=& - \frac{1}{a_S} + \frac{1}{2}\,r_S\,k^2
\nonumber \\ &+&
\sum_{n=2}^{\infty} v_{n,S}\,k^{2n}
\, , \\ \label{eq:ERE-short}
k\,\cot{\delta^{C}}\,C^2(\eta) + 2\frac{h(\eta) }{a_B}
&=& - \frac{1}{a_C} + \frac{1}{2}\,r_C\,k^2
\nonumber \\ &+&
\sum_{n=2}^{\infty} v_{n,C}\,k^{2n} \, , \label{eq:ERE-coulomb}
\end{eqnarray}
meaning that, with the exception of
the scattering length, which explicitly depends on the regularization scale 
$r_c$, see Eq.~(\ref{eq:aS-aC-contact}),
the strong and Coulomb effective range parameters for pp scattering are equal
in the present approximation
\begin{eqnarray}
r_S = r_C \, \mbox{ and } \, v_{n,S} = v_{n,C} \quad \mbox{for $n \geq 2$}\, ,
\end{eqnarray}
where $r_{S(C)}$ is the effective range, $v_{2,S(C)}$ the shape parameter, etc.
If we compare the previous results with the parameters obtained with the
Nijmegen II potential~\cite{Stoks:1994wp}, we observe a small
discrepancy~\footnote{Instead of the Nijmegen II values, it is also
possible to use the well-established experimental value
for the Coulomb pp effective range $r_C = 2.794(14)\,{\rm fm}$,
and the model dependent strong pp one $r_S = 2.84(4)\,{\rm fm}$
(see Ref.~\cite{Miller:1990iz}), although the conclusions do not
change appreciably.
}
\begin{eqnarray}
r_S = 2.84\,{\rm fm} \quad , \quad r_C = 2.76\,{\rm fm} \, ,
\end{eqnarray}
giving a $3\%$ relative difference between the strong and Coulomb parameters.
According to the error estimation of the previous section, we should expect
a relative error of $R_S / a_B$, with $R_S$ the range of the strong pp
interaction, given by one pion exchange, $R_S = R_{\pi^0} = 1/m_{\pi^0}$,
where $m_{\pi^0}$ the neutral pion mass, yielding the result 
$R_{\pi^0} / a_B = M_p\,\alpha / 2 m_{\pi^0} \sim 2.5\%$
($M_p$ is the proton mass) in agreement with the previous discrepancy
\footnote{The contributions to the strong-Coulomb effective range difference
from vacuum polarization~\cite{Heller:1960zz}, 
or from the modified Coulomb potential of Ref.~\cite{PhysRevLett.50.2039},
are expected to be much smaller than the strong (pionic) corrections.
The magnetic moment interaction~\cite{PhysRevC.42.1235} does not contribute
for s-wave proton-proton scattering.
}.

\begin{figure}[htb]
\begin{center}
\epsfig{figure=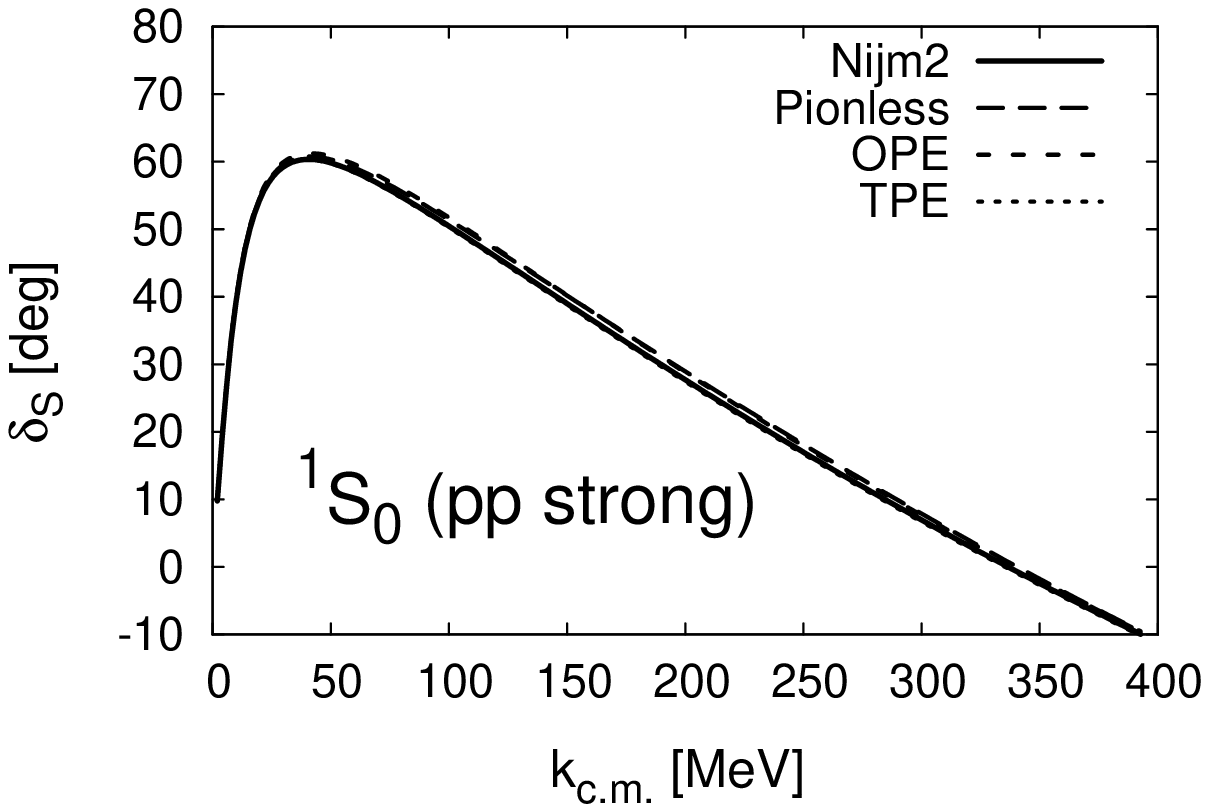, 
	height=6.0cm, width=7.0cm}
\epsfig{figure=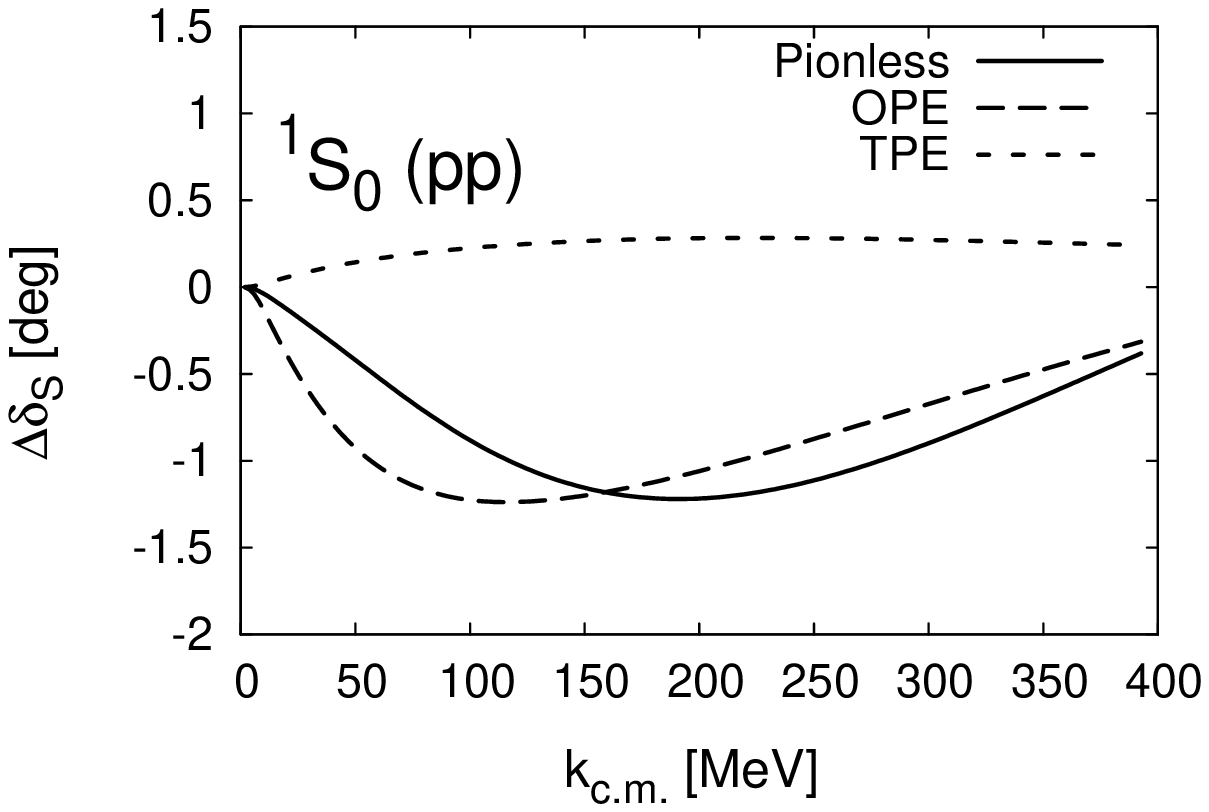, 
	height=6.0cm, width=7.0cm}
\end{center}
\caption{(Upper panel) Strong pp phase shifts computed from
the Coulomb pp phase shifts 
(Nijmegen II potential~\cite{Stoks:1994wp}) by using the zero range
strong-Coulomb correlation of Eq.~(\ref{eq:coulomb-contact}) and its
corresponding extensions when including the OPE and chiral TPE
(${\rm N^2 LO}$) potentials with a cut-off radius of $r_c = 0.1\,{\rm fm}$. 
(Lower panel) Difference between the Nijmegen phase 
shifts and those obtained with the strong-Coulomb correlations.
}
\label{fig:SC-pp}
\end{figure}

The corresponding results for the strong pp phase shifts, obtained from the
Coulomb pp ones for the Nijmegen II potential~\cite{Stoks:1994wp} by means
of Eq.~(\ref{eq:coulomb-contact}), are shown in Fig.~\ref{fig:SC-pp}.
The agreement between the strong pp Nijmegen II phase shifts and the expected
ones computed from the contact theory correlation~(\ref{eq:coulomb-contact})
is quite good, never exceeding a $1.5^{\circ}$  difference,
as expected from the relative error estimation.

\subsection{Comparison with other approaches}

The above result may be relevant to the effective field theory (EFT)
formulation of low-energy pp scattering done by Kong and
Ravndal~\cite{Kong:1998sx,Kong:1999sf} based on the power divergence
subtraction (PDS) regularization scheme of
Refs.~\cite{Kaplan:1998tg,Kaplan:1998sz}. The admitted intricacy of
the momentum space formalism in those works contrasts with the much
shorter and transparent discussion of the coordinate space
renormalization presented above.
In particular, Eq.~(\ref{eq:coulomb-contact}) implies that there are no
Coulomb corrections to the effective range once the cut-off is removed,
which is in agreement with the next-to-leading order calculation of 
Kong and Ravndal~\cite{Kong:1999sf},
but disagrees with the next-to-next-to-leading
order results of Ref.~\cite{Ando:2007fh}.
It is also implied in our treatment
that the pionless treatment of pp scattering can be made
almost scale independent if, apart from the usual counterterms 
$C_0 + C_2 (p^2 + {p'}^2) + {\mathcal O}(p^4)$, a counterterm contribution
proportional to $\alpha$ is included in the computations, 
i.e. $\alpha\,D_{e^2}$.
This observation is closely related with the results of 
Ref.~\cite{Gegelia:2003ta}, where the necessity of a strong and Coulomb
version of $C_0$ was discussed.
The previous $D_{e^2}$ counterterm fixes the difference between
the strong and Coulomb scattering length,
so the price to pay in order to remove the log
scale dependence in Eq.~(\ref{eq:aS-aC-contact}) is the impossibility
to relate the two, as both scattering lengths become input
parameters.
This seems to be in contradiction with
Kong and Ravndal~\cite{Kong:1998sx,Kong:1999sf},
who argue that the $C_2$ counterterm stabilizes the scattering length
(see also related discussions in
Refs.~\cite{Walzl:2000cx,Ando:2007fh,Ando:2008jb}).
This counterterm is analogous to the $m_{\pi}^2\,D_2$ counterterm needed
to renormalize Weinberg power counting
at leading order~\cite{Kaplan:1996xu,Beane:2001bc}.
They are both due to the similar behaviour of the Coulomb and Yukawa
potential at short distances.

Of course, these conclusions are based on our coordinate space
analysis with cut-off regularization.  
Dimensional regularization with PDS yields different 
results~\cite{Ando:2007fh}, as Coulomb corrections to the effective range
appear at next-to-next-to-leading order.
These corrections depend on the off-shell behaviour of the $\mathcal{O}(p^4)$
counterterms, which in the cut-off approach seems to be under control as long
as non-localities and off-shell ambiguities happen below $r_c$.
A more pessimistic view is presented by Gegelia in Ref.~\cite{Gegelia:2003ta},
where it is argued from the renormalization group behaviour of the counterterms
in PDS that it needs to be a strong and Coulomb version of each counterterm
(to absorb the log-divergences),
meaning that in the end it is impossible to relate
strong and Coulomb observables.
On the contrary, the renormalization group analysis with cut-off regularization
of Birse and Barford~\cite{Barford:2002je} seems to support the idea that
the Coulomb log-divergences can be absorbed in just one counterterm~\footnote{
The results of Ref.~\cite{Barford:2002je} does not exclude the existence of
Coulomb corrections to all counterterms, neither do our results
if supplemented by additional subtractions. It is just that they
are not needed in order to have scale independent results.
}.
The observations of Gegelia~\cite{Gegelia:2003ta}
can be considered as an extension to any scattering observable of the results
of Refs.~\cite{PhysRevLett.32.626,Sauer:1977hv} about the difficulty of
obtaining model independent strong scattering lengths from Coulomb ones
due to short range ambiguities of the wave function.
The previous seem to be in contradiction with usual requirement
of short distance independence of physical results in effective field theory.
In fact, as was shown in Ref.~\cite{PhysRevC.27.917},
further constraints on the short range ambiguities not considered
in~\cite{PhysRevLett.32.626,Sauer:1977hv} can noticeably reduce
the model dependence of strong parameters, in a better agreement with
EFT expectations.
Finally, we should also stress that we are only trying to separate strong
from {\it Coulomb} corrections in non-relativistic quantum mechanics.
A complete formulation on the separation of strong and {\it electromagnetic}
effects is only possible in the context of quantum field theory,
see Ref.~\cite{Gasser:2003hk} for a modern discussion on the subject.

We should nonetheless remember that cut-off regularization is a physical
regularization, in the sense that the cut-off $r_c$ can be interpreted as a
physical scale. From this point of view, the meaning of the log-divergence
in the relationship between the strong and Coulomb scattering length is
straightforward: it represents the expected error of the strong
scattering length in the pionless approximation, 
which scales as $\log(R_S/a_B)$ (instead of $R_S/a_B$, 
as in the other parameters), yielding a very large, 
about $\sim 350\%$, expected relative error
(to be compared with the one for the effective range $\sim 2.5 \%$).
One can also argue that the extra counterterm $D_{e^2}$ is not needed,
as the inclusion of the higher order components of the potential
will reduce the scale dependence.

\subsection{Strong-Coulomb Correlations with Chiral TPE Potentials}

The two potential formula makes it possible to obtain the
(experimentally inaccessible) strong pp phase shifts from the
(experimentally accessible) Coulomb ones. While a complete analysis
should of course include vacuum polarization~\cite{Heller:1960zz},
modifications to the Coulomb potential~\cite{PhysRevLett.50.2039} and
even $2 \pi \gamma $ exchange~\cite{Kaiser:2006ws,Kaiser:2006ck,Kaiser:2006na},
the interesting issue is that one can obtain model independent strong
phase shifts, provided that we employ a model independent strong pp
potential $V_S$ and model independent Coulomb phase shifts. 
Here, we will do so with an eye put on the relevant scales in the problem,
an aspect which our two-potential treatment can address
in a rather clean way. 

For the previous purpose we use the potentials of chiral perturbation
theory~\cite{Rentmeester:1999vw} as the short distance potential $V_S$
of Eq.~(\ref{eq:2V-formula-coulomb}).
These potentials include TPE
effects and can be expressed as a expansion in powers of $Q$
\begin{eqnarray}
V_S(r) = V_{\chi}^{(0)}(r) + V_{\chi}^{(2)}(r) + V_{\chi}^{(3)}(r) + 
{\mathcal O}(Q^4) \, ,
\end{eqnarray}
where $Q$ represents either the pion mass or the momentum of the
protons. We also use the Nijmegen PWA~\cite{Stoks:1993tb}, which is a
model-independent extraction of the pp s-wave {\it Coulomb}~\footnote{
It is important to notice that the pp phase shifts in the Nijmegen PWA are
not Coulomb phase shifts, but {\it electromagnetic} phase shifts.
By that it is meant that the pp phase shifts are defined with respect to the
asymptotic solutions of the full electromagnetic potential used in their
analysis, which consists on improved Coulomb, vacuum polarization and
the magnetic moment interaction (see Ref.~\cite{Stoks:1993tb} for details).
As our current analysis is not intended to be {\it complete},
we will ignore most of these details and simply 
assume that the long range potential is the usual Coulomb potential,
and that the full electromagnetic phase shifts roughly coincide with
the Coulomb ones, $\delta^{\rm EM}_{\rm PWA} \simeq \delta^{C}_{\rm PWA}$.
}
phase shifts from a large proton-proton scattering database.
With that information, we can obtain the strong pp phase shift and its
error from the PWA 
\begin{eqnarray}
\delta^C_{\rm PWA}(k) \pm \Delta \delta^C_{\rm PWA}
\rightarrow \left( V_{\chi}, r_c \right) \rightarrow
\delta^S(k) \pm \Delta \delta^S \, ,
\end{eqnarray}
and analyze the resulting cut-off dependence, which is an important
issue, as for large coordinate space cut-offs the higher order pieces
of the chiral potential are not resolved.

It should be noted here that a complete model independent separation
between strong and electromagnetic contributions is not always possible,
specially if short distance electromagnetic effects are included.
One example is the inclusion of nucleon form factor corrections 
to the magnetic moment interaction in the proton-proton PWA of
Ref.~\cite{PhysRevC.41.1435}.
Another example is proton finite size corrections to the Coulomb potential.
The formalism presented here clearly separates between {\it what we define}
as the strong and electromagnetic {\it potential}.
That does not necessarily mean that exact model independence has been achieved,
specially if corrections like the ones mentioned above are added, or
that strong and electromagnetic effects have been actually separated,
specially as electromagnetic corrections to the proton mass or to the
coupling constants have not been included.

The specific procedure we will apply is analogous to the one followed
in the pionless case, i.e., we do not directly use the strong-Coulomb
two-potential formula, Eq.~(\ref{eq:2V-formula-coulomb}), but rather
perform a subtraction of the equivalent two-potential formula for the
scattering lengths, and then check for cut-off independence of the results.
This choice also allows for a better comparison between the pionless
correlation given in Eq.~(\ref{eq:coulomb-contact}) and the corresponding
improvements when the strong physics are included explicitly.

In the present calculation we are only going to consider the chiral
potentials up to the $Q^3$ order, i.e. next-to-next-to-leading-order
(${\rm N}^2 {\rm LO}$).
At this order the finite range piece of the chiral potential consists
of one-pion exchange and chiral two-pion exchange.
An interesting feature of the chiral two-pion exchange potentials is that they
are highly singular, diverging as $\sim  1/r^6$
at ${\rm N}^2 {\rm LO}$~\footnote{The most singular (non-contact) piece of
$V_{\chi}^{(\nu)}$ will behave as $1/r^{3 + \nu}$ in coordinate space 
and as $|\vec{q}|^{\nu} f(|\vec{q}|/m_{\pi})$ in momentum space, 
being $\vec{q}$ the momentum exchanged
between the nucleons and $f$ a non-polynomial function.
}.
In harmony with previous findings~\cite{PavonValderrama:2005gu,Valderrama:2005wv,cordon:054002},
this divergence will become rather unimportant: the two potential formula
shows a smooth cut-off dependence for singular chiral potentials~\footnote{
In fact, the singular chiral two-pion exchange potentials yield smoother
results than the OPE potential. In the current regularization scheme,
OPE shows a mild log-divergence at distances of $10^{-3}\,{\rm fm}$.
This divergence can be eliminated by using a different, and more complex,
subtraction prescription, 
but for the purposes of the present discussion it is not
particularly important what happens at such small scales.
}.
In any case, we stress that our main concern is to analyze the minimal
short distance cut-off $r_c$ for which higher order effects
can be distinguished from lower order ones, rather than
the specific cut-off dependence of the results.

\begin{figure}[ttt]
\begin{center}
\epsfig{figure=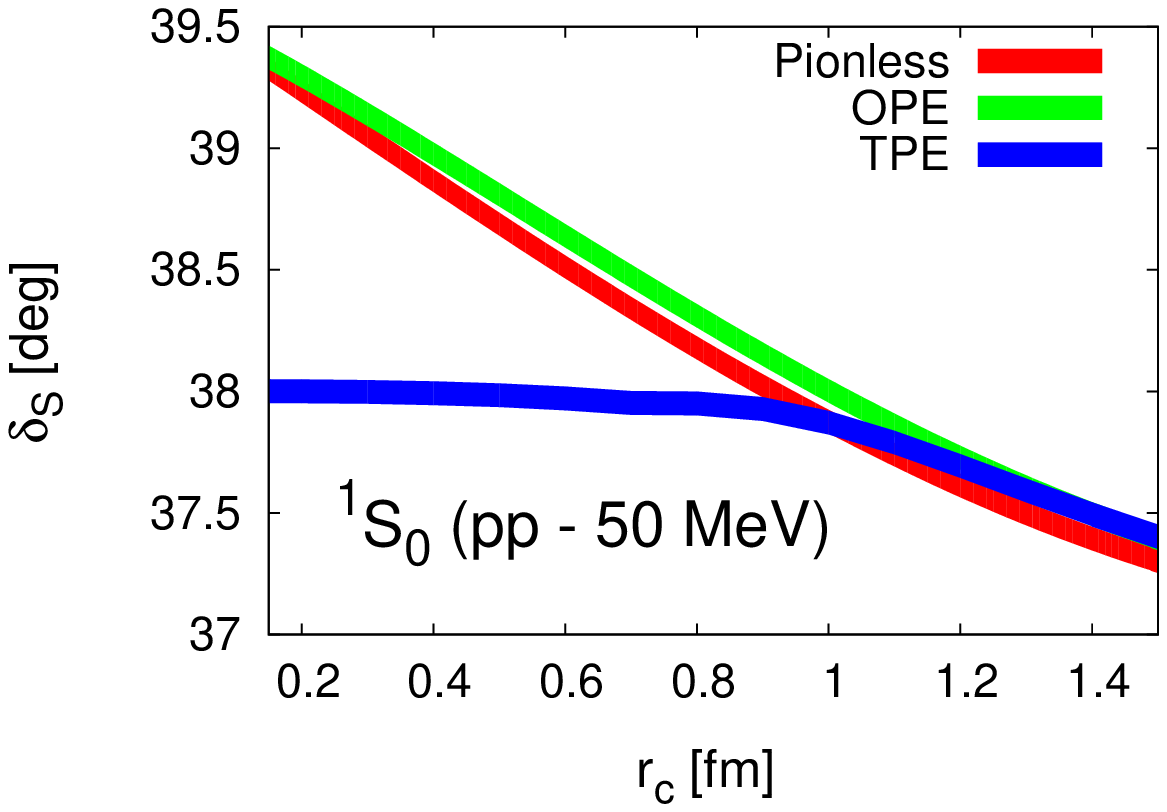, 
	height=6.0cm, width=7.0cm}
\epsfig{figure=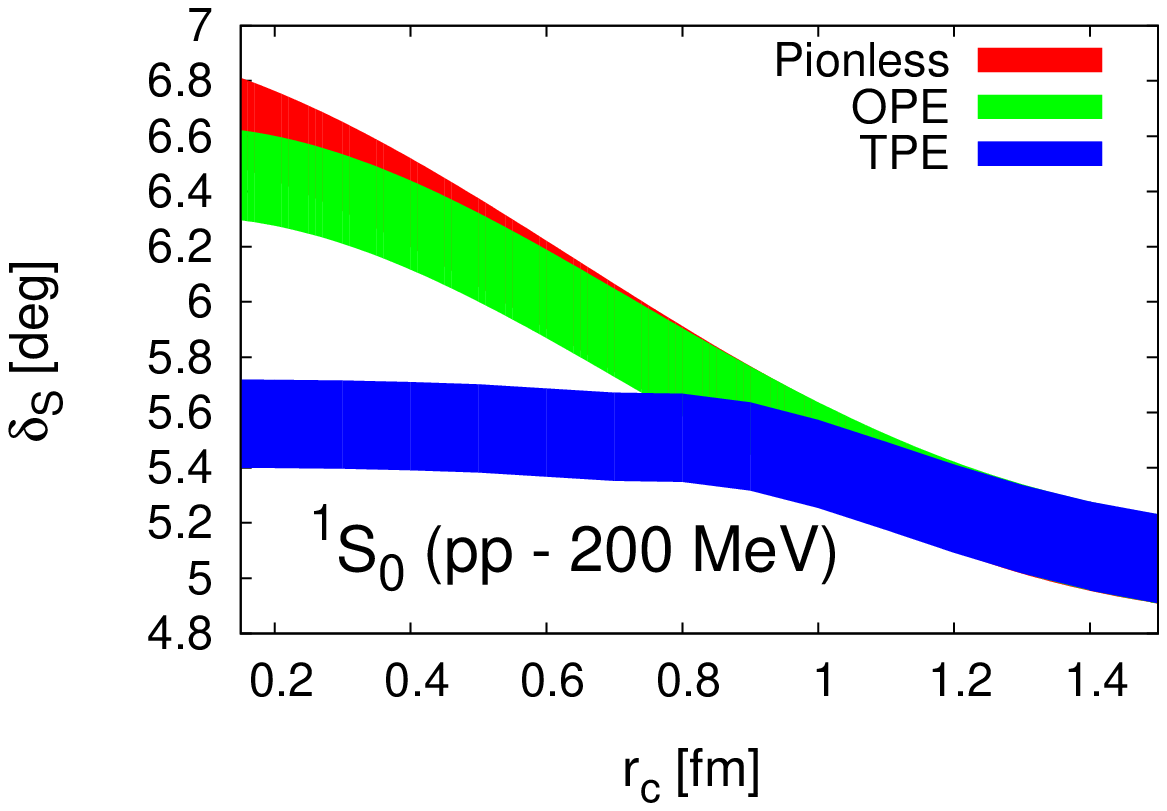, 
	height=6.0cm, width=7.0cm}
\end{center}
\caption{Strong pp s-wave phase shifts at $E_{\rm lab} = 50\,{\rm MeV}$ and
$E_{\rm lab} = 200\,{\rm MeV}$ as a function of the cut-off $r_c$, 
computed from the Coulomb pp phase shifts (Nijmegen PWA~\cite{Stoks:1993tb})
by using the chiral potential truncated at different orders in the chiral
expansion (pionless means no potential, OPE is the leading order potential
and TPE the full ${\rm N^2 LO}$ chiral potential). 
The strong and Coulomb phase shifts have been related by 
making use of Eq.~(\ref{eq:2V-formula-coulomb}) with one
subtraction at $k = 0$, in order to have
finite results when the cut-off is removed.
}
\label{fig:SC-pp-error}
\end{figure}

The results for  $E_{\rm lab} = 50\,{\rm MeV}$ and
$E_{\rm lab} = 200\,{\rm MeV}$ can be seen in Fig.~\ref{fig:SC-pp-error}.
By TPE we refer to the ${\rm N^2 LO}$ chiral potential, and for compactness
we skip the ${\rm NLO}$ results, which lie between the ${\rm LO}$ (OPE) and
the ${\rm N^2 LO}$ results.
The bands represent the error coming from the original pp Coulomb phase shift
in the PWA of Ref.~\cite{Stoks:1993tb}.
For the strong and Coulomb scattering lengths, which are needed for
the subtractions, we take the values corresponding to the 
Nijmegen II potential~\cite{Stoks:1994wp},
i.e. $\alpha_C = -7.81\,{\rm fm}$ and $\alpha_S = -17.25\,{\rm fm}$.
The actual expressions for the chiral pp potential
are taken from Ref.~\cite{Rentmeester:1999vw}.
Only the long range piece of the chiral potentials is considered, and
the corresponding counterterms are ignored: they are equivalent to a boundary
condition for the Schr\"odinger
equation~\cite{vanKolck:1998bw,vanKolck:1999mw},
and are therefore already implicitly included in the two-potential formula.
We take $g_A = 1.29$, $m_{\pi^0} = 134.98\,{\rm MeV}$ and
$f_{\pi} = 92.4\,{\rm MeV}$, which according to the definitions
of Ref.~\cite{Stoks:1993tb}
gives an $f_{pp \pi^0}^2 = 0.0755$
for the scaling mass $m_s = m_{\pi^{\pm}} = 139.57\,{\rm MeV}$.
The previous chiral pp potential explicitly depend on three chiral couplings,
$c_1$, $c_3$ and $c_4$, which appear at $\mathcal{O}(Q^3)$ in the expansion
of the potential, and which relate nucleon-nucleon and nucleon-pion scattering.
We take for these chiral couplings the values obtained
in Ref.~\cite{Rentmeester:2003mf} from analyzing the pp data alone,
i.e. $c_1 = -0.76(7)\,{\rm GeV}^{-1}$, $c_3 = -4.78(11)\,{\rm GeV}^{-1}$, 
and $c_4 = 3.92(52)\,{\rm GeV}^{-1}$.
As can be seen, for a cut-off above $r_c = 1.2\,{\rm fm}$, one cannot
distinguish, within uncertainties, between lower and higher order
computations, i.e. it does not matter whether pions are included
in $V_S$ or not.
Actually cut-offs below $r_c = 0.8\,{\rm fm}$ are needed
in order to fully distinguish the chiral two-pion exchange
contributions within the accuracy of the phase shifts. 
This result is not entirely surprising as could be anticipated from
considering the two-pion exchange related Compton wavelength scale
$\lambda_{2\pi} = 1/ 2 m_{\pi} \sim 0.7\,{\rm fm}$. 
A more complete analysis should include vacuum polarization,
$2\pi\gamma$ and $\gamma \gamma$ exchange effects,
which will affect the the precise values
of the strong phase shifts but will hardly change the observation on the
relevant scales. 
The same remarks also apply to the error analysis, which should include 
the error in the subtracted strong and Coulomb scattering lengths
and the theoretical uncertainties in the chiral potential itself,
like, for example, the error in the determination of the chiral couplings.

We can also compare the extracted strong effective ranges 
for the different cases considered.
In this case, it is used the Coulomb pp phase shift from the Nijmegen II
potential as input for the two-potential formula and the resulting
strong phase shifts are shown in Fig.~\ref{fig:SC-pp}.
For the regularization scale $r_c = 0.1\,{\rm fm}$, we obtain
\begin{eqnarray}
r_{S,{\rm contact}} &=& 2.78\,{\rm fm}\, , \\
r_{S,{\rm OPE}} &=& 2.63\,{\rm fm}\, , \\
r_{S,{\rm TPE}} &=& 2.87\,{\rm fm}\, ,
\end{eqnarray}
to be compared with the Nijmegen II result, $r_{S} = 2.84\,{\rm fm}$.
The pionless value differs from the one given in
Section~\ref{subsec:pionless}
due to finite cut-off effects, while OPE surprisingly contributes
in the wrong direction.
The TPE result reproduces the Nijmegen II one within a $1\%$ accuracy level,
and agrees within error estimations with the extraction of 
Ref.~\cite{Miller:1990iz}, $r_S = 2.84(4)\,{\rm fm}$,
where the error accounts for different sources of model-dependence.

Finally, we note that {\it despite} the TPE potential becomes highly
singular at short distances, diverging as $\sim 1/r^6$, nothing
dramatic happens, making the limit $r_c \to 0 $ innocuous {\it
  precisely} when the TPE effects become visible, i.e. for $r_c \le
0.8 {\rm fm}$~\footnote{Actually, despite Eq.~(\ref{eq:condition-short}) being
fulfilled to the extreme, the TPE correction is small because the
typical minimal wavelength is still not comparable to the range where 
the TPE correction takes over.}.
This particular feature is a specific merit of our two potential formula
which provides a clean separation between scales and implements in an
extended distorted wave fashion the renormalization program carried
out in previous works~(see e.g. \cite{PavonValderrama:2007nu,Entem:2007jg}).

\section{Conclusions}

The two potential formalism provides a framework where forces of
different origin and ranges may be disentangled rather explicitly.
We have proposed a coordinate space formulation which re-states the result
in a rather transparent way and fully exploits the boundary value character
as well as the superposition principle of the scattering problem.
Our result allows for a detailed
investigation of the relevant scales built into the problem. This is
particularly enlightening in the case of singular potentials and their
renormalization, a subject of recent interest. We have exemplified our
approach by discussing its consequences for the proton-proton system,
where the electromagnetic and strong forces contribute to the
scattering process, as a method to extract the strong phase shifts in
a model independent fashion. We have only discussed $s-$waves (with the
exception of Eq.~(\ref{eq:coulomb-contact-p-wave}))
and single channel scattering.
The extension to higher partial waves, as well as coupled channels, 
is straightforward but cumbersome, see Appendix \ref{app:extension}.
Such an extended formalism might allow to discuss further interesting
applications of the present ideas to similar problems where a scale
separation of different forces would be necessary.

\begin{acknowledgments}

We thank Evgeny Epelbaum, Jambul Gegelia, Andreas Nogga and
Daniel R. Phillips for discussions, a critical and 
careful reading of the manuscript and for pointing out several useful
references. M.P.V. is supported by the Helmholtz Association fund
provided to the young investigator group
``Few-Nucleon Systems in Chiral Effective Field Theory'' (grant
VH-NG-222), and the virtual institute ``Spin and strong QCD'' (VH-VI-231).
The work of E.R.A.  is supported in part by funds provided by the
Spanish DGI and FEDER funds with grant no. FIS2008-01143/FIS, and the
Junta de Andaluc{\'\i}a grant no. FQM225-05.  M.P.V. and E.R.A. are
supported by the EU Integrated Infrastructure Initiative Hadron
Physics Project contract no. RII3-CT-2004-506078.

\end{acknowledgments}

\appendix

\section{Extension to Higher Partial Waves and Coupled Channels}
\label{app:extension}

\subsection{Higher Partial Waves}

The extension of our two potential formula to higher partial waves
is straightforward.
The full two-body system is described by the corresponding 
reduced Schr\"odinger equation for the $l$-wave
\begin{eqnarray}
\label{eq:schroe-full-l}
-u_{k,l}'' + 
\left[ 2\mu\,(V_L(r) + V_S(r)) + \frac{l(l+1)}{r^2}\right]\,u_{k,l}(r)
&& \nonumber \\
=  k^2 u_{k,l}(r) \, , &&
\end{eqnarray}
where $u_{k,l}(r)$ is the $l$-wave reduced wave function, 
$\mu$ the reduced mass and $k$ the center-of-mass momentum.
The asymptotics of $u_{k,l}$ for $r\to\infty$ is given by
\begin{eqnarray}
\label{eq:ps-uk-l}
u_{k,l}(r) \to \cot{\delta_l}\,\hat{j}_l(k r) - \hat{y}_l(k r)\, ,
\end{eqnarray}
where $\hat{j}_l$ and ${y}_l(k r)$ are the reduced spherical Bessel functions,
defined as $\hat{j}_l(x) = x\,j_l(x)$ and $\hat{y}_l(x) = x\,y_l(x)$.
We only consider here the case of a long range potential $V_L$ decaying
faster than $1/r^2$ at large distances.
By making use of the superposition principle we rewrite the full solution as
\begin{eqnarray}
u_{k,l}(r) = \cot{\delta_l}\,J_{k,l}(r) - Y_{k,l}(r) \, ,
\end{eqnarray}
where $J_{k,l}$ and $Y_{k,l}$ are solutions of Eq.~(\ref{eq:schroe-full-l})
subjected to the asymptotic boundary conditions $J_{k,l}(r) \to \hat{j}_l(k r)$
and $Y_{k,l}(r) \to \hat{y}_l(k r)$ for $r \to \infty$.
The corresponding scattering problem for which only the short range potential
$V_S$ is present is described by the reduced Schr\"odinger equation
\begin{eqnarray}
\label{eq:schroe-short-l}
-{u_{k,l}^S}'' +
\left[ 2\mu\,V_S(r) + \frac{l(l+1)}{r^2} \right]\,u_{k,l}^S(r) 
&& \nonumber \\
= k^2 u_{k,l}^S(r) \, , &&
\end{eqnarray}
with $u_{k,l}^S$ the short $l$-wave reduced wave function.
The short distance phase shift is obtained from the asymptotic behaviour of
$u_{k,l}^S$
\begin{eqnarray}
u_{k,l}^S(r) \to \cot{\delta_l^S}\,\hat{j}_l(k r) - \hat{y}_l(k r)\, ,
\end{eqnarray}
for $r \to \infty$.
We rewrite $u_{k,l}^S$ as
\begin{eqnarray}
u_{k,l}^S(r) = \cot{\delta_l}\,J^S_{k,l}(r) - Y^S_{k,l}(r) \, ,
\end{eqnarray}
with $J^S_{k,l}$ and $Y^S_{k,l}$ solutions of Eq.~(\ref{eq:schroe-short-l})
obeying the asymptotic boundary conditions $J^S_{k,l}(r) \to \hat{j}_l(k r)$
and $Y^S_{k,l}(r) \to \hat{y}_l(k r)$.

As usual we match the logarithmic derivatives of $u_{k,l}(r)$
and $u_{k,l}^S(r)$ at the cut-off radius $r = r_c$, yielding
\begin{eqnarray}
\label{eq:2V-formula-l}
\cot \delta_l (k) = 
\frac{{\cal A}_l (k,r_c) \cot \delta_l^S (k) - {\cal B}_l (k,r_c) }
{{\cal C}_l (k,r_c) \cot \delta_l^S (k) - {\cal D}_l (k,r_c) } \, ,
\end{eqnarray} 
where ${\cal A}_l$, ${\cal B}_l$, ${\cal C}_l$ and ${\cal D}_l$ are defined as
\begin{eqnarray}
{\cal A}_l (k,r_c) &=& W(J_{k,l}^S, Y_{k,l})\big|_{r = r_c} \, , \\
{\cal B}_l (k,r_c) &=& W(Y_{k,l}^S, Y_{k,l})\big|_{r = r_c} \, , \\
{\cal C}_l (k,r_c) &=& W(J_{k,l}^S, J_{k,l})\big|_{r = r_c} \, , \\
{\cal D}_l (k,r_c) &=& W(Y_{k,l}^S, J_{k,l})\big|_{r = r_c} \, ,
\end{eqnarray}
in analogy with the $s$-wave case.
In principle the use of the previous formula is straightforward as long as a
finite cut-off is employed in the computation.
On the contrary, if one tries to remove the cut-off,
some divergences may appear, mostly related to the centrifugal barrier.
Therefore a detailed analytical or numerical study of the divergences
will be necessary in order to obtain a finite result in the $r_c \to 0$ limit.

\subsection{Coupled Channels}

The extension to the coupled channel case is direct to obtain
if an adequate notation is used.
We will consider the general case of N coupled channels.
They can be described by the following Schr\"odinger equation,
which in compact notation reads
\begin{eqnarray}
\label{eq:schroe-full-N}
-{\bf u}_{k}'' + 
\left[ 2\mu\, ({\bf V}_L(r) + {\bf V}_S(r)) + 
\frac{{\bf L}^2}{r^2}\right]\,{\bf u}_{k}(r) && \nonumber \\
=  k^2 {\bf u}_{k}(r) \, , &&
\end{eqnarray}
where the wave function ${\bf u}_k$ is now an ${\rm N} \times {\rm N}$ matrix,
each column representing a linearly independent solution.
The long and short range potentials ${\bf V}_L$ and ${\bf V}_S$ are 
also ${\rm N} \times {\rm N}$ matrices (the non-diagonal terms relating
the different channels), and ${\bf L}^2$ is the angular momentum matrix, 
which is diagonal and given by
\begin{eqnarray}
{\bf L}^2 = {\rm diag}(l_1 (l_1 + 1), l_2 (l_2 + 1), \dots, l_N (l_N + 1)) \, ,
\end{eqnarray}
being $l_1$, $l_2$, ..., $l_N$ the orbital angular momentum of each channel.
In principle there are $2 {\rm N}$ linearly independent solutions 
(two per channel), but regularity conditions at the origin reduce
this number to ${\rm N}$.
This is why the wave function can be represented by an 
${\rm N} \times {\rm N}$ matrix.
We have also added the simplifying assumption that there are no inelasticities,
meaning that in Eq.~(\ref{eq:schroe-full-N}) the source of the coupling is
either tensor forces or dipole-dipole interactions.

The asymptotic behaviour of the wave function matrix ${\bf u}_k$
is given by the following expression
\begin{eqnarray}
{\bf u}_k (r) \to {\bf j}(k r)\,{\bf M}(k) - {\bf y}(k r) \, ,
\end{eqnarray}
for $r \to \infty$, where ${\bf j}$ and ${\bf y}$ are diagonal matrices
given by
\begin{eqnarray}
{\bf j}(k r) &=& {\rm diag}(\hat{j}_{l_1}(k r), \hat{j}_{l_2}(k r), \dots,
\hat{j}_{l_N}(k r)) \, , \\
{\bf y}(k r) &=& {\rm diag}(\hat{y}_{l_1}(k r), \hat{y}_{l_2}(k r), \dots,
\hat{y}_{l_N}(k r)) \, ,
\end{eqnarray}
with $\hat{j}_l$ and $\hat{y}_l$ the reduced spherical Bessel functions
as defined in the previous section.
The matrix ${\bf M}(k)$ is the analogous of $\cot{\delta}$ for coupled
channels and is related to the $S$-matrix by 
${\bf M}(k) = i\,({\bf S}(k) + {\bf 1})/({\bf S}(k) - {\bf 1})$
with ${\bf 1}$ the identity matrix.
It is a symmetric matrix and contains $N(N+1)/2$ independent scattering
parameters or {\it phase shifts}.
By making use of the superposition principle, we can rewrite the wave 
function matrix as~\footnote{
The reason why we write ${\bf J}_k (r)\,{\bf M}(k)$ instead of 
${\bf M}(k)\,{\bf J}_k (r)$ in Eq.~(\ref{eq:u-lin-N}) 
is because
if ${\bf u}_k$ is a solution of the Schr\"odinger equation 
(\ref{eq:schroe-full-N}) and ${\bf A}$ a constant 
${\rm N} \times {\rm N}$ matrix,
then ${\bf u}_k \, {\bf A}$ is also a solution
(but this is not the case for ${\bf A}\,{\bf u}_k$).
}
\begin{eqnarray}
\label{eq:u-lin-N}
{\bf u}_k (r) = {\bf J}_k (r)\,{\bf M}(k) - {\bf Y}_k (r) \, ,
\end{eqnarray}
where ${\bf J}_k$ and ${\bf Y}_k$ are solutions of Eq.~(\ref{eq:schroe-full-N})
which asymptotically behave as ${\bf J}_k (r) \to  {\bf j}(k r)$ and
${\bf Y}_k (r) \to  {\bf y}(k r)$.

The corresponding Schr\"odinger equation for the short wave function is
\begin{eqnarray}
\label{eq:schroe-short-N}
-{{\bf u}_{k}^S}'' + 
\left[ 2\mu\,{\bf V}_S(r) + 
\frac{{\bf L}^2}{r^2}\right]\,{\bf u}_{k}^S(r) 
=  k^2 {\bf u}_{k}^S(r) \, , &&
\end{eqnarray}
where, in analogy to the full case, the short wave function matrix
can be written as
\begin{eqnarray}
\label{eq:uS-lin-N}
{\bf u}^S_k (r) = {\bf J}^S_k (r)\,{\bf M}^S(k) - {\bf Y}^S_k (r) \, ,
\end{eqnarray}
with ${\bf J}^S_k$ and ${\bf Y}^S_k$ solutions of Eq.~(\ref{eq:schroe-short-N})
subjected to the asymptotic boundary condition 
${\bf J}^S_k (r) \to  {\bf j}(k r)$ and ${\bf Y}^S_k (r) \to  {\bf y}(k r)$
for $r \to \infty$.

For obtaining the corresponding two potential formula one needs to match
the logarithmic derivatives of the wave functions,
which for the coupled channel case means
\begin{eqnarray}
\label{eq:bc-equality-N}
{{\bf u}_k}'(r_c) {({\bf u}_k(r_c))}^{-1} = 
{{\bf u}^S_k}'(r_c) {({{\bf u}^S_k}(r_c))}^{-1} \, .
\end{eqnarray}
Using the following Wronskian relation
\begin{eqnarray}
{\bf u}_k^{\rm T}(r_c) {{\bf u}_k}'(r_c)  &=&
{{\bf u}_k^{\rm T}}'(r_c) {{\bf u}_k}(r_c) \, ,
\end{eqnarray}
where the $^{\rm T}$ superscript denotes the transpose, the boundary condition
given by Eq.~(\ref{eq:bc-equality-N}) can be rewritten as
\begin{eqnarray}
{\bf u}_k^{\rm T}(r_c)\,{{\bf u}^S_k}'(r_c) =
{{\bf u}^{\rm T}_k}'(r_c)\,{{\bf u}_k^S}(r_c) \, ,
\end{eqnarray}
an expression which does not involve the inverse of the wave functions.
If we rewrite ${\bf u}_k$ and ${\bf u}_k^S$ in terms of Eq.~(\ref{eq:u-lin-N})
and (\ref{eq:uS-lin-N}), we arrive at our final expression
\begin{eqnarray}
{\bf M}(k) &=& 
\left( {\bf {\cal A}}(k, r_c)\,{\bf M}^S(k) - {\bf {\cal B}}(k, r_c)\right)
\times \nonumber \\ &&
{\left( {\bf {\cal C}}(k, r_c)\,{\bf M}^S(k) - 
{\bf {\cal D}}(k, r_c)\right)}^{-1}
\end{eqnarray}
with ${\cal A}$, ${\cal B}$, ${\cal C}$ and ${\cal D}$ defined as
\begin{eqnarray}
{\cal A} (k,r_c) &=& - W({\bf Y}_{k}^{\rm T}, {\bf J}_{k}^S)\big|_{r = r_c} \, , \\
{\cal B} (k,r_c) &=& - W({\bf Y}_{k}^{\rm T}, {\bf Y}_{k}^S)\big|_{r = r_c} \, , \\
{\cal C} (k,r_c) &=& - W({\bf J}_{k}^{\rm T}, {\bf J}_{k}^S)\big|_{r = r_c} \, , \\
{\cal D} (k,r_c) &=& - W({\bf J}_{k}^{\rm T}, {\bf Y}_{k}^S)\big|_{r = r_c} \, ,
\end{eqnarray}
where the Wronskian is given by $W({\bf F}, {\bf G})|_{r_c} = 
{\bf F}'(r_c) {\bf G}(r_c) - {\bf F}(r_c) {\bf G}'(r_c)$.


\begin{thebibliography}{46}
\expandafter\ifx\csname natexlab\endcsname\relax\def\natexlab#1{#1}\fi
\expandafter\ifx\csname bibnamefont\endcsname\relax
  \def\bibnamefont#1{#1}\fi
\expandafter\ifx\csname bibfnamefont\endcsname\relax
  \def\bibfnamefont#1{#1}\fi
\expandafter\ifx\csname citenamefont\endcsname\relax
  \def\citenamefont#1{#1}\fi
\expandafter\ifx\csname url\endcsname\relax
  \def\url#1{\texttt{#1}}\fi
\expandafter\ifx\csname urlprefix\endcsname\relax\def\urlprefix{URL }\fi
\providecommand{\bibinfo}[2]{#2}
\providecommand{\eprint}[2][]{\url{#2}}

\bibitem[{\citenamefont{Watson}(1952)}]{PhysRev.88.1163}
\bibinfo{author}{\bibfnamefont{K.~M.} \bibnamefont{Watson}},
  \bibinfo{journal}{Phys. Rev.} \textbf{\bibinfo{volume}{88}},
  \bibinfo{pages}{1163} (\bibinfo{year}{1952}).

\bibitem[{\citenamefont{Gell-Mann and Goldberger}(1953)}]{PhysRev.91.398}
\bibinfo{author}{\bibfnamefont{M.}~\bibnamefont{Gell-Mann}} \bibnamefont{and}
  \bibinfo{author}{\bibfnamefont{M.~L.} \bibnamefont{Goldberger}},
  \bibinfo{journal}{Phys. Rev.} \textbf{\bibinfo{volume}{91}},
  \bibinfo{pages}{398} (\bibinfo{year}{1953}).

\bibitem[{\citenamefont{Entem et~al.}(2008)\citenamefont{Entem, Ruiz~Arriola,
  Pavon~Valderrama, and Machleidt}}]{Entem:2007jg}
\bibinfo{author}{\bibfnamefont{D.~R.} \bibnamefont{Entem}},
  \bibinfo{author}{\bibfnamefont{E.}~\bibnamefont{Ruiz~Arriola}},
  \bibinfo{author}{\bibfnamefont{M.}~\bibnamefont{Pavon~Valderrama}},
  \bibnamefont{and}
  \bibinfo{author}{\bibfnamefont{R.}~\bibnamefont{Machleidt}},
  \bibinfo{journal}{Phys. Rev.} \textbf{\bibinfo{volume}{C77}},
  \bibinfo{pages}{044006} (\bibinfo{year}{2008}), \eprint{0709.2770}.

\bibitem[{\citenamefont{Pavon~Valderrama and
  Arriola}(2008)}]{PavonValderrama:2007nu}
\bibinfo{author}{\bibfnamefont{M.}~\bibnamefont{Pavon~Valderrama}}
  \bibnamefont{and} \bibinfo{author}{\bibfnamefont{E.~R.}
  \bibnamefont{Arriola}}, \bibinfo{journal}{Annals Phys.}
  \textbf{\bibinfo{volume}{323}}, \bibinfo{pages}{1037} (\bibinfo{year}{2008}),
  \eprint{0705.2952}.

\bibitem[{\citenamefont{Landau and Smorodinsky}(1944)}]{Landau44}
\bibinfo{author}{\bibnamefont{Landau}} \bibnamefont{and}
  \bibinfo{author}{\bibnamefont{Smorodinsky}}, \bibinfo{journal}{J. Phys. Acad.
  Sci. U.S.S.R.} \textbf{\bibinfo{volume}{8}} (\bibinfo{year}{1944}).

\bibitem[{\citenamefont{Holstein}(1999)}]{Holstein:1999nq}
\bibinfo{author}{\bibfnamefont{B.~R.} \bibnamefont{Holstein}},
  \bibinfo{journal}{Phys. Rev.} \textbf{\bibinfo{volume}{D60}},
  \bibinfo{pages}{114030} (\bibinfo{year}{1999}), \eprint{nucl-th/9901041}.

\bibitem[{\citenamefont{Pavon~Valderrama and
  Ruiz~Arriola}(2005)}]{PavonValderrama:2005gu}
\bibinfo{author}{\bibfnamefont{M.}~\bibnamefont{Pavon~Valderrama}}
  \bibnamefont{and}
  \bibinfo{author}{\bibfnamefont{E.}~\bibnamefont{Ruiz~Arriola}},
  \bibinfo{journal}{Phys. Rev.} \textbf{\bibinfo{volume}{C72}},
  \bibinfo{pages}{054002} (\bibinfo{year}{2005}), \eprint{nucl-th/0504067}.

\bibitem[{\citenamefont{Pavon~Valderrama and
  Arriola}(2006)}]{Valderrama:2005wv}
\bibinfo{author}{\bibfnamefont{M.}~\bibnamefont{Pavon~Valderrama}}
  \bibnamefont{and} \bibinfo{author}{\bibfnamefont{E.~R.}
  \bibnamefont{Arriola}}, \bibinfo{journal}{Phys. Rev.}
  \textbf{\bibinfo{volume}{C74}}, \bibinfo{pages}{054001}
  (\bibinfo{year}{2006}), \eprint{nucl-th/0506047}.

\bibitem[{\citenamefont{Cord\'{o}n and Arriola}(2008)}]{cordon:054002}
\bibinfo{author}{\bibfnamefont{A.~C.} \bibnamefont{Cord\'{o}n}}
  \bibnamefont{and} \bibinfo{author}{\bibfnamefont{E.~R.}
  \bibnamefont{Arriola}}, \bibinfo{journal}{Physical Review C (Nuclear
  Physics)} \textbf{\bibinfo{volume}{78}}, \bibinfo{eid}{054002}
  (pages~\bibinfo{numpages}{17}) (\bibinfo{year}{2008}).

\bibitem[{\citenamefont{Abramowitz and Stegun}(1964)}]{abramowitz+stegun}
\bibinfo{author}{\bibfnamefont{M.}~\bibnamefont{Abramowitz}} \bibnamefont{and}
  \bibinfo{author}{\bibfnamefont{I.~A.} \bibnamefont{Stegun}},
  \emph{\bibinfo{title}{Handbook of Mathematical Functions with Formulas,
  Graphs, and Mathematical Tables}} (\bibinfo{publisher}{Dover},
  \bibinfo{address}{New York}, \bibinfo{year}{1964}), \bibinfo{edition}{ninth
  dover printing, tenth gpo printing} ed., ISBN \bibinfo{isbn}{0-486-61272-4}.

\bibitem[{\citenamefont{Kok et~al.}(1982)\citenamefont{Kok, de~Maag, Brouwer,
  and van Haeringen}}]{Kok:1982dr}
\bibinfo{author}{\bibfnamefont{L.~P.} \bibnamefont{Kok}},
  \bibinfo{author}{\bibfnamefont{J.~W.} \bibnamefont{de~Maag}},
  \bibinfo{author}{\bibfnamefont{H.~H.} \bibnamefont{Brouwer}},
  \bibnamefont{and} \bibinfo{author}{\bibfnamefont{H.}~\bibnamefont{van
  Haeringen}}, \bibinfo{journal}{Phys. Rev.} \textbf{\bibinfo{volume}{C26}},
  \bibinfo{pages}{2381} (\bibinfo{year}{1982}).

\bibitem[{\citenamefont{Partensky and Ericson}(1967)}]{Partensky1967382}
\bibinfo{author}{\bibfnamefont{A.}~\bibnamefont{Partensky}} \bibnamefont{and}
  \bibinfo{author}{\bibfnamefont{M.}~\bibnamefont{Ericson}},
  \bibinfo{journal}{Nuclear Physics B} \textbf{\bibinfo{volume}{1}},
  \bibinfo{pages}{382 } (\bibinfo{year}{1967}), ISSN \bibinfo{issn}{0550-3213}.

\bibitem[{\citenamefont{de~Maag et~al.}(1984)\citenamefont{de~Maag, Kok, and
  van Haeringen}}]{deMaag:1983nz}
\bibinfo{author}{\bibfnamefont{J.~W.} \bibnamefont{de~Maag}},
  \bibinfo{author}{\bibfnamefont{L.~P.} \bibnamefont{Kok}}, \bibnamefont{and}
  \bibinfo{author}{\bibfnamefont{H.}~\bibnamefont{van Haeringen}},
  \bibinfo{journal}{J. Math. Phys.} \textbf{\bibinfo{volume}{25}},
  \bibinfo{pages}{684} (\bibinfo{year}{1984}).

\bibitem[{\citenamefont{Blatt and Jackson}(1949)}]{Blatt49}
\bibinfo{author}{\bibfnamefont{J.~M.} \bibnamefont{Blatt}} \bibnamefont{and}
  \bibinfo{author}{\bibfnamefont{J.~D.} \bibnamefont{Jackson}},
  \bibinfo{journal}{Phys. Rev.} \textbf{\bibinfo{volume}{76}},
  \bibinfo{pages}{18} (\bibinfo{year}{1949}).

\bibitem[{\citenamefont{Jackson and Blatt}(1950)}]{RevModPhys.22.77}
\bibinfo{author}{\bibfnamefont{J.~D.} \bibnamefont{Jackson}} \bibnamefont{and}
  \bibinfo{author}{\bibfnamefont{J.~M.} \bibnamefont{Blatt}},
  \bibinfo{journal}{Rev. Mod. Phys.} \textbf{\bibinfo{volume}{22}},
  \bibinfo{pages}{77} (\bibinfo{year}{1950}).

\bibitem[{\citenamefont{Bertulani et~al.}(2002)\citenamefont{Bertulani, Hammer,
  and Van~Kolck}}]{Bertulani:2002sz}
\bibinfo{author}{\bibfnamefont{C.~A.} \bibnamefont{Bertulani}},
  \bibinfo{author}{\bibfnamefont{H.~W.} \bibnamefont{Hammer}},
  \bibnamefont{and}
  \bibinfo{author}{\bibfnamefont{U.}~\bibnamefont{Van~Kolck}},
  \bibinfo{journal}{Nucl. Phys.} \textbf{\bibinfo{volume}{A712}},
  \bibinfo{pages}{37} (\bibinfo{year}{2002}), \eprint{nucl-th/0205063}.

\bibitem[{\citenamefont{Bethe}(1949)}]{PhysRev.76.38}
\bibinfo{author}{\bibfnamefont{H.~A.} \bibnamefont{Bethe}},
  \bibinfo{journal}{Phys. Rev.} \textbf{\bibinfo{volume}{76}},
  \bibinfo{pages}{38} (\bibinfo{year}{1949}).

\bibitem[{\citenamefont{Stoks et~al.}(1994)\citenamefont{Stoks, Klomp,
  Terheggen, and de~Swart}}]{Stoks:1994wp}
\bibinfo{author}{\bibfnamefont{V.~G.~J.} \bibnamefont{Stoks}},
  \bibinfo{author}{\bibfnamefont{R.~A.~M.} \bibnamefont{Klomp}},
  \bibinfo{author}{\bibfnamefont{C.~P.~F.} \bibnamefont{Terheggen}},
  \bibnamefont{and} \bibinfo{author}{\bibfnamefont{J.~J.}
  \bibnamefont{de~Swart}}, \bibinfo{journal}{Phys. Rev.}
  \textbf{\bibinfo{volume}{C49}}, \bibinfo{pages}{2950} (\bibinfo{year}{1994}),
  \eprint{nucl-th/9406039}.

\bibitem[{\citenamefont{Miller et~al.}(1990)\citenamefont{Miller, Nefkens, and
  Slaus}}]{Miller:1990iz}
\bibinfo{author}{\bibfnamefont{G.~A.} \bibnamefont{Miller}},
  \bibinfo{author}{\bibfnamefont{B.~M.~K.} \bibnamefont{Nefkens}},
  \bibnamefont{and} \bibinfo{author}{\bibfnamefont{I.}~\bibnamefont{Slaus}},
  \bibinfo{journal}{Phys. Rept.} \textbf{\bibinfo{volume}{194}},
  \bibinfo{pages}{1} (\bibinfo{year}{1990}).

\bibitem[{\citenamefont{Heller}(1960)}]{Heller:1960zz}
\bibinfo{author}{\bibfnamefont{L.}~\bibnamefont{Heller}},
  \bibinfo{journal}{Phys. Rev.} \textbf{\bibinfo{volume}{120}},
  \bibinfo{pages}{627} (\bibinfo{year}{1960}).

\bibitem[{\citenamefont{Austen and de~Swart}(1983)}]{PhysRevLett.50.2039}
\bibinfo{author}{\bibfnamefont{G.~J.~M.} \bibnamefont{Austen}}
  \bibnamefont{and} \bibinfo{author}{\bibfnamefont{J.~J.}
  \bibnamefont{de~Swart}}, \bibinfo{journal}{Phys. Rev. Lett.}
  \textbf{\bibinfo{volume}{50}}, \bibinfo{pages}{2039} (\bibinfo{year}{1983}).

\bibitem[{\citenamefont{Stoks and de~Swart}(1990)}]{PhysRevC.42.1235}
\bibinfo{author}{\bibfnamefont{V.~G.~J.} \bibnamefont{Stoks}} \bibnamefont{and}
  \bibinfo{author}{\bibfnamefont{J.~J.} \bibnamefont{de~Swart}},
  \bibinfo{journal}{Phys. Rev. C} \textbf{\bibinfo{volume}{42}},
  \bibinfo{pages}{1235} (\bibinfo{year}{1990}).

\bibitem[{\citenamefont{Kong and Ravndal}(1999)}]{Kong:1998sx}
\bibinfo{author}{\bibfnamefont{X.}~\bibnamefont{Kong}} \bibnamefont{and}
  \bibinfo{author}{\bibfnamefont{F.}~\bibnamefont{Ravndal}},
  \bibinfo{journal}{Phys. Lett.} \textbf{\bibinfo{volume}{B450}},
  \bibinfo{pages}{320} (\bibinfo{year}{1999}), \eprint{nucl-th/9811076}.

\bibitem[{\citenamefont{Kong and Ravndal}(2000)}]{Kong:1999sf}
\bibinfo{author}{\bibfnamefont{X.}~\bibnamefont{Kong}} \bibnamefont{and}
  \bibinfo{author}{\bibfnamefont{F.}~\bibnamefont{Ravndal}},
  \bibinfo{journal}{Nucl. Phys.} \textbf{\bibinfo{volume}{A665}},
  \bibinfo{pages}{137} (\bibinfo{year}{2000}), \eprint{hep-ph/9903523}.

\bibitem[{\citenamefont{Kaplan et~al.}(1998)\citenamefont{Kaplan, Savage, and
  Wise}}]{Kaplan:1998tg}
\bibinfo{author}{\bibfnamefont{D.~B.} \bibnamefont{Kaplan}},
  \bibinfo{author}{\bibfnamefont{M.~J.} \bibnamefont{Savage}},
  \bibnamefont{and} \bibinfo{author}{\bibfnamefont{M.~B.} \bibnamefont{Wise}},
  \bibinfo{journal}{Phys. Lett.} \textbf{\bibinfo{volume}{B424}},
  \bibinfo{pages}{390} (\bibinfo{year}{1998}), \eprint{nucl-th/9801034}.

\bibitem[{\citenamefont{Kaplan et~al.}(1999)\citenamefont{Kaplan, Savage, and
  Wise}}]{Kaplan:1998sz}
\bibinfo{author}{\bibfnamefont{D.~B.} \bibnamefont{Kaplan}},
  \bibinfo{author}{\bibfnamefont{M.~J.} \bibnamefont{Savage}},
  \bibnamefont{and} \bibinfo{author}{\bibfnamefont{M.~B.} \bibnamefont{Wise}},
  \bibinfo{journal}{Phys. Rev.} \textbf{\bibinfo{volume}{C59}},
  \bibinfo{pages}{617} (\bibinfo{year}{1999}), \eprint{nucl-th/9804032}.

\bibitem[{\citenamefont{Ando et~al.}(2007)\citenamefont{Ando, Shin, Hyun, and
  Hong}}]{Ando:2007fh}
\bibinfo{author}{\bibfnamefont{S.-i.} \bibnamefont{Ando}},
  \bibinfo{author}{\bibfnamefont{J.~W.} \bibnamefont{Shin}},
  \bibinfo{author}{\bibfnamefont{C.~H.} \bibnamefont{Hyun}}, \bibnamefont{and}
  \bibinfo{author}{\bibfnamefont{S.~W.} \bibnamefont{Hong}},
  \bibinfo{journal}{Phys. Rev.} \textbf{\bibinfo{volume}{C76}},
  \bibinfo{pages}{064001} (\bibinfo{year}{2007}), \eprint{0704.2312}.

\bibitem[{\citenamefont{Gegelia}(2004)}]{Gegelia:2003ta}
\bibinfo{author}{\bibfnamefont{J.}~\bibnamefont{Gegelia}},
  \bibinfo{journal}{Eur. Phys. J.} \textbf{\bibinfo{volume}{A19}},
  \bibinfo{pages}{355} (\bibinfo{year}{2004}), \eprint{nucl-th/0310012}.

\bibitem[{\citenamefont{Walzl et~al.}(2001)\citenamefont{Walzl, Meissner, and
  Epelbaum}}]{Walzl:2000cx}
\bibinfo{author}{\bibfnamefont{M.}~\bibnamefont{Walzl}},
  \bibinfo{author}{\bibfnamefont{U.~G.} \bibnamefont{Meissner}},
  \bibnamefont{and} \bibinfo{author}{\bibfnamefont{E.}~\bibnamefont{Epelbaum}},
  \bibinfo{journal}{Nucl. Phys.} \textbf{\bibinfo{volume}{A693}},
  \bibinfo{pages}{663} (\bibinfo{year}{2001}), \eprint{nucl-th/0010019}.

\bibitem[{\citenamefont{Ando and Birse}(2008)}]{Ando:2008jb}
\bibinfo{author}{\bibfnamefont{S.-i.} \bibnamefont{Ando}} \bibnamefont{and}
  \bibinfo{author}{\bibfnamefont{M.~C.} \bibnamefont{Birse}},
  \bibinfo{journal}{Phys. Rev.} \textbf{\bibinfo{volume}{C78}},
  \bibinfo{pages}{024004} (\bibinfo{year}{2008}), \eprint{0805.3655}.

\bibitem[{\citenamefont{Kaplan et~al.}(1996)\citenamefont{Kaplan, Savage, and
  Wise}}]{Kaplan:1996xu}
\bibinfo{author}{\bibfnamefont{D.~B.} \bibnamefont{Kaplan}},
  \bibinfo{author}{\bibfnamefont{M.~J.} \bibnamefont{Savage}},
  \bibnamefont{and} \bibinfo{author}{\bibfnamefont{M.~B.} \bibnamefont{Wise}},
  \bibinfo{journal}{Nucl. Phys.} \textbf{\bibinfo{volume}{B478}},
  \bibinfo{pages}{629} (\bibinfo{year}{1996}), \eprint{nucl-th/9605002}.

\bibitem[{\citenamefont{Beane et~al.}(2002)\citenamefont{Beane, Bedaque,
  Savage, and van Kolck}}]{Beane:2001bc}
\bibinfo{author}{\bibfnamefont{S.~R.} \bibnamefont{Beane}},
  \bibinfo{author}{\bibfnamefont{P.~F.} \bibnamefont{Bedaque}},
  \bibinfo{author}{\bibfnamefont{M.~J.} \bibnamefont{Savage}},
  \bibnamefont{and} \bibinfo{author}{\bibfnamefont{U.}~\bibnamefont{van
  Kolck}}, \bibinfo{journal}{Nucl. Phys.} \textbf{\bibinfo{volume}{A700}},
  \bibinfo{pages}{377} (\bibinfo{year}{2002}), \eprint{nucl-th/0104030}.

\bibitem[{\citenamefont{Barford and Birse}(2003)}]{Barford:2002je}
\bibinfo{author}{\bibfnamefont{T.}~\bibnamefont{Barford}} \bibnamefont{and}
  \bibinfo{author}{\bibfnamefont{M.~C.} \bibnamefont{Birse}},
  \bibinfo{journal}{Phys. Rev.} \textbf{\bibinfo{volume}{C67}},
  \bibinfo{pages}{064006} (\bibinfo{year}{2003}), \eprint{hep-ph/0206146}.

\bibitem[{\citenamefont{Sauer}(1974)}]{PhysRevLett.32.626}
\bibinfo{author}{\bibfnamefont{P.~U.} \bibnamefont{Sauer}},
  \bibinfo{journal}{Phys. Rev. Lett.} \textbf{\bibinfo{volume}{32}},
  \bibinfo{pages}{626} (\bibinfo{year}{1974}).

\bibitem[{\citenamefont{Sauer and Walliser}(1977)}]{Sauer:1977hv}
\bibinfo{author}{\bibfnamefont{P.~U.} \bibnamefont{Sauer}} \bibnamefont{and}
  \bibinfo{author}{\bibfnamefont{H.}~\bibnamefont{Walliser}},
  \bibinfo{journal}{J. Phys.} \textbf{\bibinfo{volume}{G3}},
  \bibinfo{pages}{1513} (\bibinfo{year}{1977}).

\bibitem[{\citenamefont{Rahman and Miller}(1983)}]{PhysRevC.27.917}
\bibinfo{author}{\bibfnamefont{M.}~\bibnamefont{Rahman}} \bibnamefont{and}
  \bibinfo{author}{\bibfnamefont{G.~A.} \bibnamefont{Miller}},
  \bibinfo{journal}{Phys. Rev. C} \textbf{\bibinfo{volume}{27}},
  \bibinfo{pages}{917} (\bibinfo{year}{1983}).

\bibitem[{\citenamefont{Gasser et~al.}(2003)\citenamefont{Gasser, Rusetsky, and
  Scimemi}}]{Gasser:2003hk}
\bibinfo{author}{\bibfnamefont{J.}~\bibnamefont{Gasser}},
  \bibinfo{author}{\bibfnamefont{A.}~\bibnamefont{Rusetsky}}, \bibnamefont{and}
  \bibinfo{author}{\bibfnamefont{I.}~\bibnamefont{Scimemi}},
  \bibinfo{journal}{Eur. Phys. J.} \textbf{\bibinfo{volume}{C32}},
  \bibinfo{pages}{97} (\bibinfo{year}{2003}), \eprint{hep-ph/0305260}.

\bibitem[{\citenamefont{Kaiser}(2006{\natexlab{a}})}]{Kaiser:2006ws}
\bibinfo{author}{\bibfnamefont{N.}~\bibnamefont{Kaiser}},
  \bibinfo{journal}{Phys. Rev.} \textbf{\bibinfo{volume}{C73}},
  \bibinfo{pages}{044001} (\bibinfo{year}{2006}{\natexlab{a}}),
  \eprint{nucl-th/0601099}.

\bibitem[{\citenamefont{Kaiser}(2006{\natexlab{b}})}]{Kaiser:2006ck}
\bibinfo{author}{\bibfnamefont{N.}~\bibnamefont{Kaiser}},
  \bibinfo{journal}{Phys. Rev.} \textbf{\bibinfo{volume}{C73}},
  \bibinfo{pages}{064003} (\bibinfo{year}{2006}{\natexlab{b}}),
  \eprint{nucl-th/0605040}.

\bibitem[{\citenamefont{Kaiser}(2006{\natexlab{c}})}]{Kaiser:2006na}
\bibinfo{author}{\bibfnamefont{N.}~\bibnamefont{Kaiser}},
  \bibinfo{journal}{Phys. Rev.} \textbf{\bibinfo{volume}{C74}},
  \bibinfo{pages}{067001} (\bibinfo{year}{2006}{\natexlab{c}}),
  \eprint{nucl-th/0610089}.

\bibitem[{\citenamefont{Rentmeester et~al.}(1999)\citenamefont{Rentmeester,
  Timmermans, Friar, and de~Swart}}]{Rentmeester:1999vw}
\bibinfo{author}{\bibfnamefont{M.~C.~M.} \bibnamefont{Rentmeester}},
  \bibinfo{author}{\bibfnamefont{R.~G.~E.} \bibnamefont{Timmermans}},
  \bibinfo{author}{\bibfnamefont{J.~L.} \bibnamefont{Friar}}, \bibnamefont{and}
  \bibinfo{author}{\bibfnamefont{J.~J.} \bibnamefont{de~Swart}},
  \bibinfo{journal}{Phys. Rev. Lett.} \textbf{\bibinfo{volume}{82}},
  \bibinfo{pages}{4992} (\bibinfo{year}{1999}), \eprint{nucl-th/9901054}.

\bibitem[{\citenamefont{Stoks et~al.}(1993)\citenamefont{Stoks, Kompl,
  Rentmeester, and de~Swart}}]{Stoks:1993tb}
\bibinfo{author}{\bibfnamefont{V.~G.~J.} \bibnamefont{Stoks}},
  \bibinfo{author}{\bibfnamefont{R.~A.~M.} \bibnamefont{Kompl}},
  \bibinfo{author}{\bibfnamefont{M.~C.~M.} \bibnamefont{Rentmeester}},
  \bibnamefont{and} \bibinfo{author}{\bibfnamefont{J.~J.}
  \bibnamefont{de~Swart}}, \bibinfo{journal}{Phys. Rev.}
  \textbf{\bibinfo{volume}{C48}}, \bibinfo{pages}{792} (\bibinfo{year}{1993}).

\bibitem[{\citenamefont{Bergervoet et~al.}(1990)\citenamefont{Bergervoet, van
  Campen, Klomp, de~Kok, Rijken, Stoks, and de~Swart}}]{PhysRevC.41.1435}
\bibinfo{author}{\bibfnamefont{J.~R.} \bibnamefont{Bergervoet}},
  \bibinfo{author}{\bibfnamefont{P.~C.} \bibnamefont{van Campen}},
  \bibinfo{author}{\bibfnamefont{R.~A.~M.} \bibnamefont{Klomp}},
  \bibinfo{author}{\bibfnamefont{J.-L.} \bibnamefont{de~Kok}},
  \bibinfo{author}{\bibfnamefont{T.~A.} \bibnamefont{Rijken}},
  \bibinfo{author}{\bibfnamefont{V.~G.~J.} \bibnamefont{Stoks}},
  \bibnamefont{and} \bibinfo{author}{\bibfnamefont{J.~J.}
  \bibnamefont{de~Swart}}, \bibinfo{journal}{Phys. Rev. C}
  \textbf{\bibinfo{volume}{41}}, \bibinfo{pages}{1435} (\bibinfo{year}{1990}).

\bibitem[{\citenamefont{van Kolck}(1999{\natexlab{a}})}]{vanKolck:1998bw}
\bibinfo{author}{\bibfnamefont{U.}~\bibnamefont{van Kolck}},
  \bibinfo{journal}{Nucl. Phys.} \textbf{\bibinfo{volume}{A645}},
  \bibinfo{pages}{273} (\bibinfo{year}{1999}{\natexlab{a}}),
  \eprint{nucl-th/9808007}.

\bibitem[{\citenamefont{van Kolck}(1999{\natexlab{b}})}]{vanKolck:1999mw}
\bibinfo{author}{\bibfnamefont{U.}~\bibnamefont{van Kolck}},
  \bibinfo{journal}{Prog. Part. Nucl. Phys.} \textbf{\bibinfo{volume}{43}},
  \bibinfo{pages}{337} (\bibinfo{year}{1999}{\natexlab{b}}),
  \eprint{nucl-th/9902015}.

\bibitem[{\citenamefont{Rentmeester et~al.}(2003)\citenamefont{Rentmeester,
  Timmermans, and de~Swart}}]{Rentmeester:2003mf}
\bibinfo{author}{\bibfnamefont{M.~C.~M.} \bibnamefont{Rentmeester}},
  \bibinfo{author}{\bibfnamefont{R.~G.~E.} \bibnamefont{Timmermans}},
  \bibnamefont{and} \bibinfo{author}{\bibfnamefont{J.~J.}
  \bibnamefont{de~Swart}}, \bibinfo{journal}{Phys. Rev.}
  \textbf{\bibinfo{volume}{C67}}, \bibinfo{pages}{044001}
  (\bibinfo{year}{2003}), \eprint{nucl-th/0302080}.

\end{thebibliography}

\end{document}